\documentclass{elsart}
\usepackage{epsfig}
\begin{document}
\begin{frontmatter}
\title{Microscopic foundations of the Rayleigh law of hysteresis}
\author[INFM]{Stefano Zapperi\corauthref{cor1}}
\corauth[cor1]{Corresponding author. Phone: +390649913437
Fax:+39064463158} \ead{zapperi@pil.phys.uniroma1.it}
\author[IEN]{Alessandro Magni}
\author[IEN]{Gianfranco Durin}

\address[INFM]{INFM sezione di Roma 1, Dipartimento di Fisica, Universit\`a "La
Sapienza", P.le A. Moro 2, 00185 Roma, Italy}
\address[IEN]{Istituto Elettrotecnico Nazionale Galileo Ferraris and
INFM, Corso M. d'Azeglio 42, I-10125 Torino, Italy}

\begin{abstract}
The hysteresis properties of ferromagnetic materials at low
field are described by the Rayleigh law. We analyze the problem
in the light of modern statistical mechanics models of hysteresis.
In particular, we compute the demagnetization curve and
derive the Rayleigh parameters $a$ and $b$
in the random-field Ising model and in a model of
domain wall depinning. In the random-field Ising model
the Rayleigh law is obeyed only in the disorder dominated phase,
while in the low disorder phase it is not possible to demagnetize
the sample. This approach allows us to link $a$ and $b$ to microstructural
parameters, such as the domain wall energy,
the internal disorder or the exchange interactions. Finally,
our results are compared with experiments.
\end{abstract}
\end{frontmatter}

\section{Introduction}

In 1887 Lord Rayleigh analyzed the hysteresis properties of ferromagnetic
materials at low fields, close to the demagnetized state
\cite{RAY-87,Bertotti,Chikazumi}.
When the field is cycled between $-H_m$ and $H_m$ ,
the magnetization $M$ is found to follow
a simple quadratic law $M = (a+bH_{m})\pm b(H_{m}^2-H^2)/2$, where
the  signs $\pm$ distinguish the upper and lower branch of the loop.
The area of the loop can easily be computed and is given by
$W=(4/3) b H_m^3$. In addition, the response of the system to
a small field change, starting from the demagnetized
state is given by $M_m=aH_m\pm b H_m^2$. The Rayleigh law
has been shown to hold in several ferromagnetic materials \cite{VER-81}
and it has been also widely applied to describe ferroelectric
materials \cite{DAM-97,BOL-00}.
A few papers have reported significant deviations from the simple
quadratic law but no explanation has been provided \cite{BER-91}.

In 1942 the Rayleigh law was interpreted by N\'eel in terms of the
motion of a point (i.e. a rigid domain wall) in a random energy
landscape, whose statistical properties determine the value of $a$
and $b$ \cite{NEE-42}. In particular, $a$ is associated to
reversible motions inside one of the many minima of the random
potential, while $b$ describes irreversible jumps between
different valleys. Several generalizations of this approach have
been proposed, considering more refined forms for the random
energy landscape \cite{PFE-67,KRO-70,KRO-79,KRO-92,MAG-99}. These approaches
assume a single rigid domain wall and thus do not consider the
effect of domain wall bowing and the interactions between
different domains. When this effects are important it is hard to
link the Rayleigh law to  the microstructural properties of the material, 
such as dimensionality, chemical composition or interaction types. 

Here, we reconsider all these problems in light of novel
statistical mechanics approaches to hysteresis. In particular we compute
the Rayleigh parameters $a$ and $b$ in the random-field Ising model
\cite{SET-93} and in a model of domain wall depinning
\cite{HIL-76,NAT-90,ENO-94,CHU-95,URB-95,CIZ-97,ZAP-98}.
This approach allows us to link $a$ and $b$
to microstructural parameters, such as the domain wall
energy, the internal disorder or the exchange interactions.

\section{Models}

A microscopic theory of hysteresis should in principle
recover the Rayleigh law from the collective properties of
interacting magnetic moments. The Rayleigh parameters would then
be expressed in terms of micromagnetic parameters, that
are available in the literature for a variety of ferromagnetic
materials. Ferromagnets can in general be described by a
locally varying magnetization $M_i(\vec{r})$ evolving under the action
of the externally applied field. The evolution of $M_i$ is ruled
by appropriate relaxation equations that can be written in terms
of the micromagnetic energy \cite{Bertotti,Chikazumi}.
Considering for simplicity a uniaxial material, we can write the energy as
\begin{equation}
E=\sum_{i=1}^{3}\int d^3r  [A(\vec{\nabla}M_i)^2+K(M_i n_i)^2-
\mu_0 M_s(H_i+H_{dem}^{(i)})M_i],
\label{eq:tot}
\end{equation}
where $A$ is the exchange interaction,  $K$ is the anisotropy
constant, $n_i$ is the anisotropy axis,
and $H_{dem}^{(i)}$ is the component $i$ of the demagnetizing
field and $M_s$ is the saturation magnetization.

An essential contribution to the properties of the hysteresis
loops is given by disorder, due to crystal imperfections, internal
stresses, non magnetic impurities, that are present in most
magnetic materials. N\'eel recognized this fact and replaced the
micromagnetic free energy by a random function of the
magnetization (i.e. $E=E_R(M)-HM$) \cite{NEE-42}. The
approximation is rather drastic but makes the problem analytically
tractable. In general we can model the disorder by quenched local
fluctuations of the micromagnetic parameters. Disordered
interaction terms  are conventionally denoted as random bonds ($A
\to A(\vec{r})$), random anisotropies ($n_i \to  n_i (\vec{r})$)
and random fields ($H \to H_{ext}+h(\vec{r})$). A complete
solution of micromagnetic equations including disorder is a very
complicated task. One should then resort to some kind of
approximation. In recent years two main approaches have been
undertaken to describe the magnetization properties of disordered
materials.

The first approach takes into account the
fact that in soft magnetic materials
demagnetizing fields give rise to broad domains and the
magnetization process is dominated by domain wall motion.
One can thus reduce the problem to the motion of a flexible
domain wall in a random potential.
The domain wall contribution to the micromagnetic
free energy can be expressed in term of
the domain wall coordinates $z(\vec{x})$, inserting in Eq.~(\ref{eq:tot})
$M_i(\vec{r})=\delta_{i3} M_s g((r_3-z(\vec{x}))/\delta_w)$
in Eq.~(\ref{eq:tot}), where $\delta_w\simeq \sqrt{A/K}$
is the domain wall width and $g(x)=\pm 1$ for $x \to \pm\infty$
\cite{HIL-76,CHU-95,ZAP-98}:
\begin{equation}
E_{dw}=\int d^2x[\gamma_w (\nabla z(\vec{x}))^2 - z(\vec{x}) \mu_0 M_s (H+
H_d(\{z(\vec{x})\})+V(\vec{x},z(\vec{x}))]
\label{eq:edw}
\end{equation}
where the domain wall energy is given by
$\gamma_w\simeq\sqrt{AK}$, the stray field is given by
$H_d=\mu_0M_s^2\int d^2x'~ \partial z/\partial x' (x-x')/|r-r'|^3$ and $V$ is a
random function taking into account all the disorder
contributions. In the following we will consider $V$ as a
superposition of pinning centers randomly distributed in space
\begin{equation}
V(\vec{x},z)=\sum_p f_0 \exp -((z-Z_p(\vec{x}))/\xi_p)^2
\end{equation}
where $f_0$ is the strength of each pinning center, of width $\xi_p$
located in $Z_p$. The effective form of the pinning potential
is not essential as long as the interaction range $\xi_p$ is 
finite. This is not the case for pinning due to isolated dislocations,
whose stress field decays as $1/r$. One should notice, however,
that dislocations are typically arranged in patterns where the
stress field is screened, providing us with an effective correlation
length. In the following, we will restrict our attention to
cases in which the pinning correlation length is smaller than 
the domain wall width, so that we can replace $\xi_p$ by $\delta_w$.
This case corresponds mainly to the effect of non-magnetic impurities
(see Ref.~\cite{NAT-90} for a discussion of this point).

Models based on Eq.~\ref{eq:edw} have been used in the past to
compute the coercive field \cite{HIL-76,ENO-94}, analyze thermal
relaxation \cite{NAT-90}, explain the statistical
properties of the Barkhausen noise
\cite{URB-95,CIZ-97,ZAP-98,DUR-00} and describe magnetization
creep in thin films \cite{LEM-98}. 

An approach based on domain wall motion is not adequate to describe the
magnetization properties of hard ferromagnets, where the
presence of strong random anisotropies prevents the formation
of extended domains. For this class of materials a description
in terms of interacting spins seems more appropriate.
Several disordered spin models have been proposed in the
past to describe hysteresis. Among those, the simplest
and most studied is the RFIM \cite{SET-93,PER-99}, where
a spin $s_i = \pm 1$ is assigned
to each site $i$ of a $d-$dimensional lattice. The spins
are coupled to their nearest-neighbors spins by a ferromagnetic
interaction of strength $J$ and to the external field $H$.
In addition, to each site of the lattice it is associated
a random field $h_i$ taken from a Gaussian probability distribution
with variance $R$, $P(h)=\exp(-h^2/2R^2)/\sqrt{2\pi}R$.
The Hamiltonian thus reads
\begin{equation}
E = -\sum_{\langle i,j \rangle}Js_i s_j -\sum_i(H+h_i)s_i,
\label{eq:rfim}
\end{equation}
where the first sum is restricted to nearest-neighbors pairs.
The dynamics proposed in Ref.~\cite{SET-93}
is such that  the spins align with the local field
\begin{equation}
s_i = \mbox{sign}(J\sum_j s_j  + h_i +H).
\end{equation}
In this way a single spin flip can lead the neighboring spins
to flip, eventually trigger an avalanche. Using this dynamics
it has been shown that the RFIM displays a phase transition
in $d\geq 3$ as a function of $R$ \cite{SET-93,PER-99}.
For $R<R_c$, the saturation loop has a discrete jump
in the magnetization at $H=\pm H_c$.
The jump disappears for $R>R_c$ and $R=R_c$ corresponds to a
critical point and the model satisfies scaling laws \cite{SET-93,PER-99}.

\section{Collective pinning effects in domain wall hysteresis}

The hysteresis properties of interfaces in random media have
been studied in the past in the context of frictional sliding \cite{BOC-97}.
Similar studies for ferromagnetic domain walls has only been
restricted to the case of high driving frequencies and large fields
\cite{LYU-99}. It is interesting to note, however, that
interfaces in random media obey return point memory, a typical
properties of ferromagnetic hysteresis.
Here, we use a model based on Eq.~(\ref{eq:edw}) to analyze
the Rayleigh law at low frequency. For the sake of simplicity
we do not consider here demagnetizing fields, which are essential,
however, to account for the large scale behavior of the magnetization
\cite{ZAP-98}. When considering small scale displacements 
of the domain wall, we expect the domain wall energy (scaling as $q^2$ 
in Fourier space a deformation of wave vector $\vec{q}$.) 
to be more relevant of the stray field contribution (scaling as $q$). 
The following arguments can nevertheless be generalized
including the effects of demagnetizing fields \cite{HIL-76}.

The scaling properties of the Rayleigh parameters can be
obtained using collective pinning theory \cite{HIL-76,NAT-90,LEM-98}.
The central concept is the identification of a coherence length
$L_c$, defining a region of space where the domain wall moves freely
from the pinning centers.  The coherence length can be obtained
comparing the domain wall energy with the disorder fluctuations
over a region of length $L_c$, considering only small transverse wall
deformations  of the order of $\delta_w$
\begin{equation}
\gamma_w \delta_w^2 \sim f_0 (n_0 L_c^2 \delta_w)^{1/2} ~~~~
L_c \sim (\gamma_w\delta_w^{3/2})/(n_0^{1/2} f_0).
\end{equation}
The depinning field $H_c$, which can be identified with
the coercive field, is obtained comparing the pinning energy
with the magnetostatic energy over a region of length $L_c$
\cite{HIL-76,NAT-90,LEM-98}
\begin{equation}
\mu_0 M_s H_c L_c^2 \delta_w \sim f_0 (n_0 L_c^2 \delta_w)^{1/2}
~~~~ H_c = (n_0 f_0^2)/(\gamma_w \delta_w^2 \mu_0 M_s).
\end{equation}
This expression for the coercive field recovers the
result obtained in Ref.~\cite{HIL-76} when two dimensional
domain vaulting is considered. In models based on a rigid
domain wall \cite{NEE-42,KRO-70,KRO-79} the coercive field is
instead proportional to the standard deviation of the pinning
field (i.e. $H_c \propto f_0 \sqrt{n_0}$).
 
Under the application of a small external field the ``unpinned''
regions of the domain wall will bow slightly and contribute to
the magnetization \cite{NAT-90}.
The expression for the susceptibility $a=dM/dH$
is similar to the one found in textbooks when discussing reversible
susceptibility \cite{Bertotti,Chikazumi}, with exception that
the bowing length is now given by $L_c$,
\begin{equation}
a\sim \mu_0 M_s^2 L_c^2/\gamma_w \sim (\gamma_w \delta_w^3)/(n_0f_0^2).
\label{eq:a}
\end{equation}
This result differs from estimates based on generalizations 
of N\'eel theory
\cite{NEE-42,KRO-70,KRO-79} where $\gamma_w$ does
not appear. 

The first correction to the linear susceptibility, and hence the
parameter $b$, can be obtained noticing that a small applied field
can in principle lead to local depinning events, thus increasing
the coherence length \cite{BOC-97}. Using the arguments reported
in Ref.~\cite{BOC-97}, one can show that, to lowest order in $H$,
$L_c(H) =L_c(0)(1+ c|H|/H_c)$ where $c$ is a numerical factor.
Inserting this expression in Eq.~(\ref{eq:a}) and expanding for
small $H$, we obtain $b \sim a/H_c$. A similar expression was
already reported by N\'eel in 1942 \cite{NEE-42}.

\begin{figure}[h]

        \epsfxsize=14cm
        \epsfbox{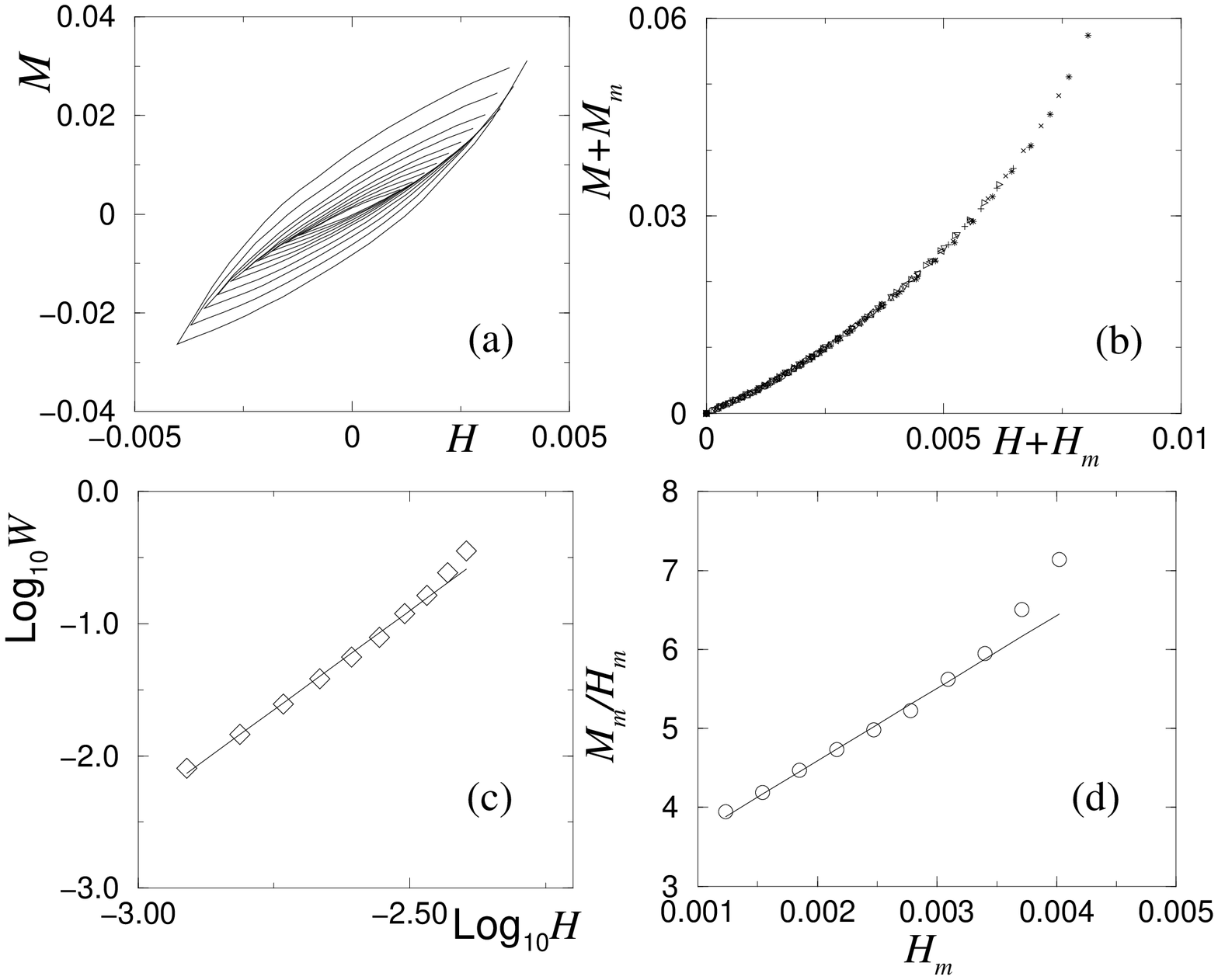}
        \vspace*{1cm}

\caption{The Rayleigh law computed in flexible domain wall in a random medium.
(a) Hysteresis loop for different values of $H_m$. (b) The lower branch
of the loops can be rescaled according to the Rayleigh law. (c) The area
of the loop scales as $W\sim H_m^3$ (the line has a slope of $3$.
(d) The Rayleigh parameter can be also obtained by a linear fit of
$M_m/H_m$ vs $H_m$. All units are arbitrary.}
\label{fig:1}
\end{figure}

\begin{figure}[h]

        \epsfxsize=14cm
        \epsfbox{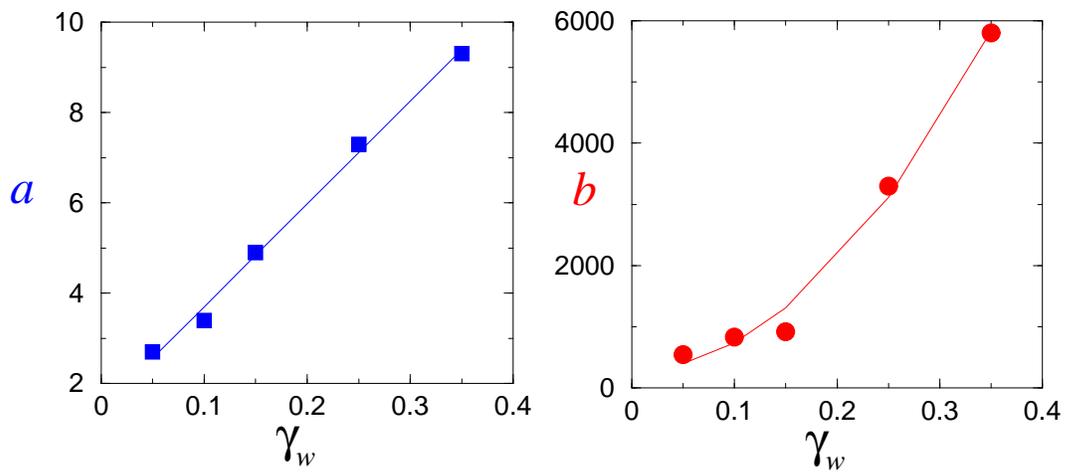}
        \vspace*{1cm}

\caption{The Rayleigh parameter $a$ and $b$ are computed for different
values of the domain wall energy $\gamma_w$. The results satisfy the relation
$a\propto \gamma_w$ and $b\propto \gamma_w^2$. All units are arbitrary.}
\label{fig:2}
\end{figure}

In order to test these consideration, we perform numerical simulations
based on Eq.~\ref{eq:edw}. The coordinates of the domain walls evolve according
to an overdamped equation $\Gamma d z/dt = -\delta E/\delta z $, where
$\Gamma$ is an effective viscosity. The equation of motion is discretized
on a grid of size $50x50$ and solved by an
adaptive-stepsize Runge-Kutta method.
We use quasistatic driving condition, applying a field $H$ and integrating
the motion until the domain wall comes to rest. The system is first
demagnetized, with the successive application of positive and negative fields
of decreasing amplitude, and then we cycle the field between $-H_m$ and
$H_m$. The Rayleigh parameters are extracted from the scaling properties
of the hysteresis loops (see Fig.~\ref{fig:1}) which are averaged
over several realizations of the disorder. The procedure is repeated
for different values of the domain wall energy $\gamma_w$ and in
Fig~\ref{fig:2} we show that simulations are in agreement with the scaling
theory.

\section{Random-field Ising model}

The RFIM is probably the simplest model showing the
combined effect of disorder and exchange interaction on the shape
of the hysteresis loop. The model can be solved exactly in one
dimension and minor loops can also be computed \cite{SHU-00}.
Recently we have been able to obtain exact results
for the entire demagnetization process in $d=1$,
including the Rayleigh laws. Here, we present numerical results
for the Rayleigh laws in $d=2$ and $d=3$.

\begin{figure}[h]

        \epsfxsize=14cm
        \epsfbox{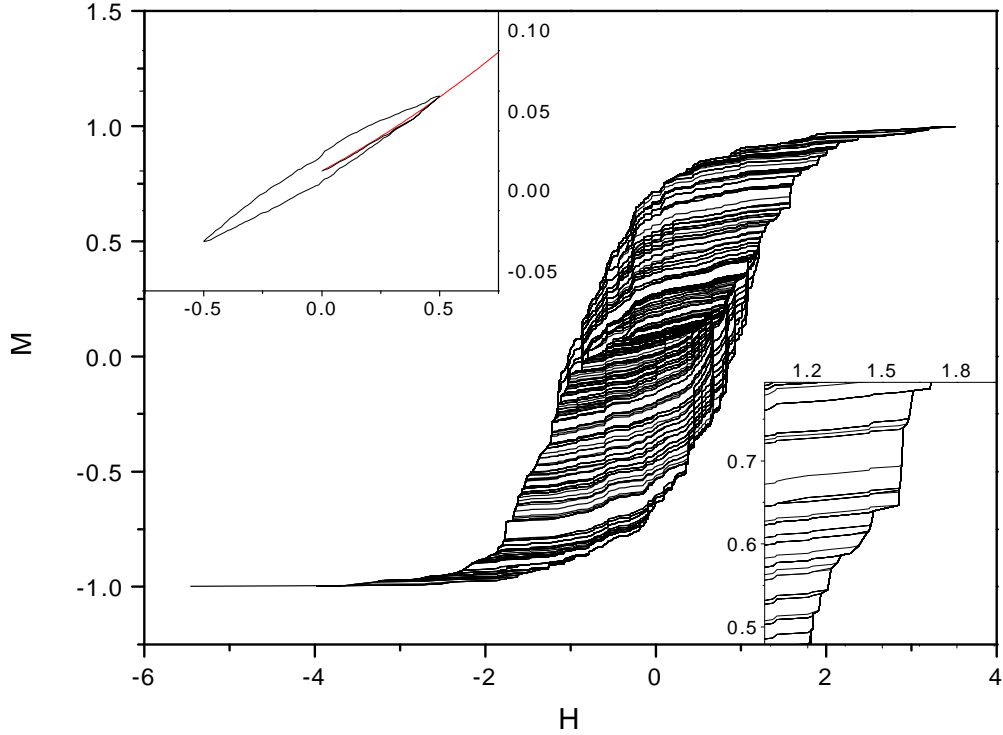}
        \vspace*{1cm}

\caption{An example of exact demagnetization process
of the RFIM in $d=2$ for $R=1.64$.  In the lower inset
we show a detail of the main figure.
In the upper inset we show the demagnetization
curve and a small loop around the demagnetized state, both
averaged over 30 realizations of the disorder. In the lower inset
we show a detail of the main figure. All units are arbitrary.}
\label{fig:3}
\end{figure}

In $d=2$, we perform a {\em perfect demagnetization} and thus
obtain unambiguously the demagnetized state for a given
realization of the disorder. This is done in practice changing the
field by precisely the amount necessary to flip the first unstable
spin \cite{PER-99}. In this way, the field is cycled between
$-H_m$ and $H_m$ and $H_m$ is then decreased at the next cycle by
precisely the amount necessary to have one avalanche less than in
the previous cycle. This corresponds to decrease $H_m$ at each
cycle by an amount $\Delta H$, with $\Delta H \to 0^+$. We thus
obtain the demagnetizing curve and extract the Rayleigh parameters
close to the demagnetized state for a system of linear size $L=50$
(see Fig.~\ref{fig:3}) for different values of $R$. During
demagnetization, in higher order loops we observe that the same
set of spins flips at precisely the same field, despite the fact
that the demagnetization proceeds and the initial state is
different.

Performing a perfect demagnetization, however,
it is not possible to go to very large system sizes. We have thus
resorted to a different algorithm which perform an {\em approximate
demagnetization}: instead of cycling the field between $-H_m$ and $H_m$
we just flip the field between these two values. The no-passing rule
\cite{SET-93} ensures that the system will go into the same
states that would be obtained decreasing and increasing adiabatically
$H$ between $-H_m$ and $H_m$. The only difference is that, after each
cycle, we have to  decrease $H_m$ by an arbitrary value $\Delta H$
(i.e. we can not perform the limit $\Delta H\to 0^+$ exactly).
We have confirmed by numerical simulations that the results
obtained with an {\em approximate demagnetization} with
$\Delta H < 10^{-3}$ are in good agreement
with the ones obtained under a {\em perfect demagnetization}.

An approximate demagnetization is then used in $d=3$, in order
to analyze the effect of the phase transition on demagnetization
and the Rayleigh law (the transition is not present in $d=1$, while
in $d=2$ the issue is controversial). We find that demagnetization
is possible only for $R>R_c$, where $R_c=2.16$ for $J=1$ \cite{PER-99}.
For $R<R_c$ the demagnetization curve coincide with the saturation loop
and it is thus not possible to define the Rayleigh parameters.
We thus measure $a$ and $b$ for different values of
$R>R_c$ and linear system sizes ranging from $L=25$ to $L=100$.
The results show that $a$ and $b$ vanish for $R\to R_c$ and
follows a scaling law $a \sim (R-R_c)^{\beta_a}$ and
$b \sim (R-R_c)^{\beta_b}$, with $\beta_a\simeq \beta_b \simeq 0.5$.
This result suggests that the demagnetization curve scales simply as
$M=(R-R_c)^{1/2}~m(H)$ for $H\to 0$. It is interesting to compare this
result with the behavior expected for $H=H_c$ on the saturation loop:
in that case the magnetization scales with an exponent $\beta\simeq 0.04$,
while $\beta=1/2$ is valid only in mean-field theory.

\section{Experimental comparison and conclusions}

In order to test the result obtained above, we perform a set of
experiments on magnetostrictive ribbons of
Fe$_{64}$Co$_{21}$B$_{15}$ amorphous alloy under moderate tensile
stress \cite{APP-99,DUR-99}. This material is characterized by
extended domain walls and the statistical properties of its
Barkhausen noise have been recently shown to be well described by
models of domain wall pinning \cite{DUR-00,DUR-99}. We thus expect
that its hysteretic behavior could also be described by collective
pinning theory.

\begin{figure}[h]

        \epsfxsize=14cm
        \epsfbox{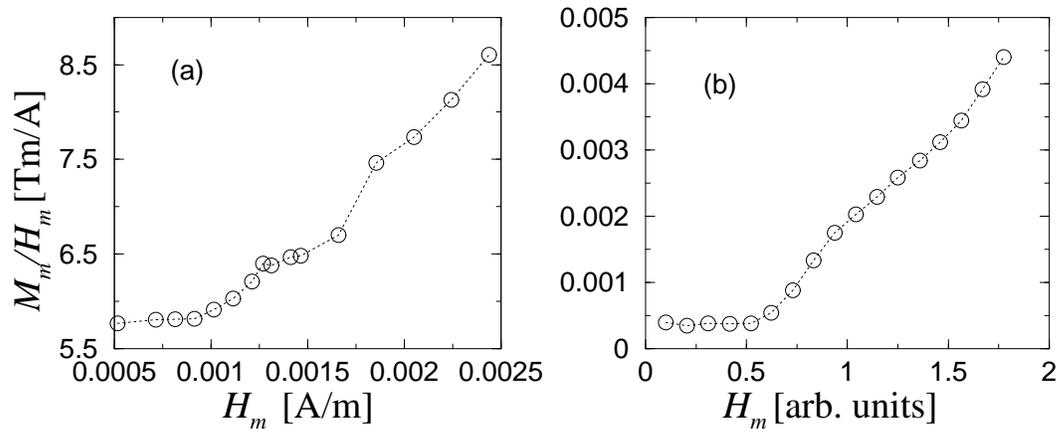}
        \vspace*{1cm}

\caption{The peak susceptibility $M_m/H_m$ as a function of $H_m$:
(a) for a single realization of the disorder in the domain wall model
and (b) for a Fe$_{64}$Co$_{21}$B$_{15}$ amorphous ribbon.}
\label{fig:5}
\end{figure}

The sample is first demagnetized and hysteresis loops are measured
at low fields. We find that the Rayleigh law is not perfectly
verified at very low fields, but the hysteresis cycle becomes
parabolic at higher field. At very low fields the magnetization is completely
reversible, then for higher fields a definite and {\it reproducible}
magnetization jump appears. Finally, at still higher fields several
jumps combined together give rise to what can be
well approximated by a parabolic cycle.
This behavior is summarized in Fig.~\ref{fig:5}a where we report
the behavior of $M_m/H_m$ as a function of $H_m$. The linear
behavior at high field is described the Rayleigh law
and the parameters $a$ and $b$ are consistent with estimates
obtained from the loop shape.

In order to understand this behavior we have simulated the
domain wall model for a {\it single realization} of the disorder
using a grid of linear size $L=100$. Also in this case, we observe
at low fields a reversible response and a jump at higher fields.
After the jump the hysteresis loop is well described by the Rayleigh
law (see Fig.~\ref{fig:5}b).
Notice that when we average the cycles over different disorder
configurations, the jump disappears. Clearly, disorder averaging can
not be performed directly for experimental data,
unless different samples are used. One should thus keep this
in mind when Rayleigh parameters are estimated.

In conclusions, we have discussed the occurrence
of the Rayleigh law of hysteresis, analyzing two classes
of models: disordered spin models and domain wall depinning
models. This is a step towards the ambitious goal to
recover hysteresis loops properties from the magnetic microstructure.


\newpage

\begin{figure}[h]

        \epsfxsize=14cm
        \epsfbox{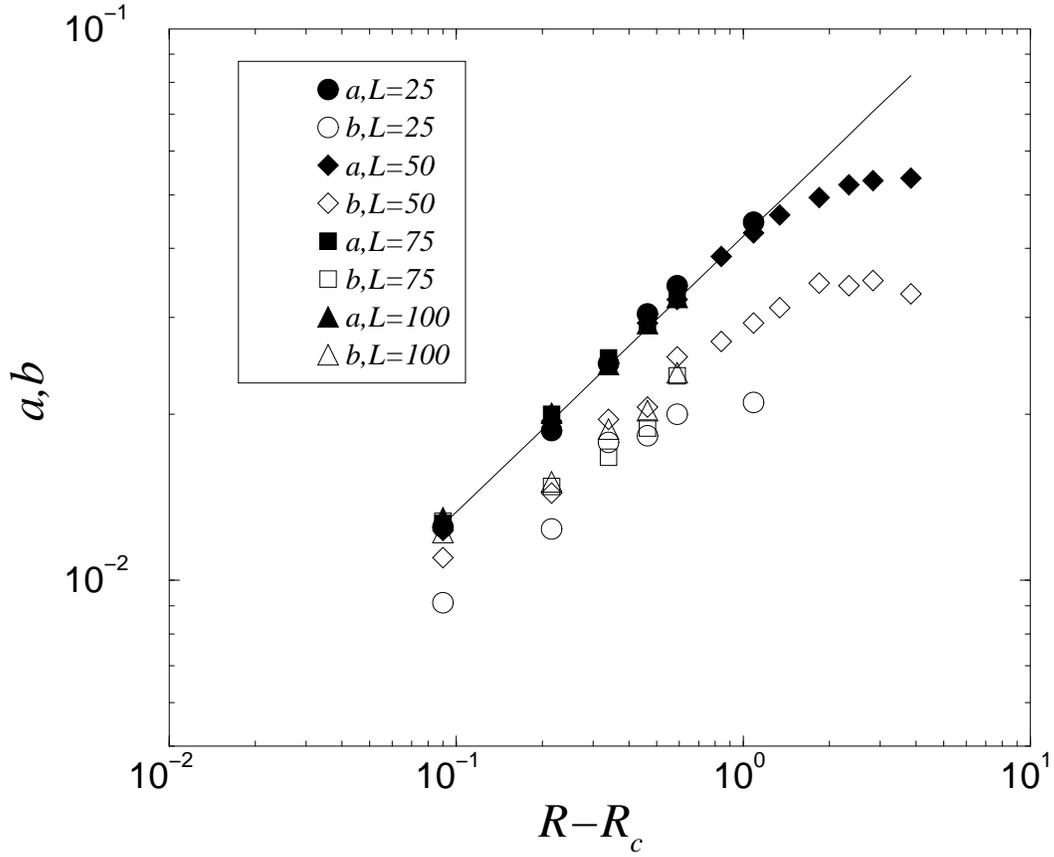}
        \vspace*{1cm}

\caption{The Rayleigh parameters $a$ and $b$ are computed for different
values of the disorder $R$ in the RFIM in $d=3$, the results are the
average over $N$ realization of the disorder: $N=100$ for $L=25$.
$N=50$ for $L=50$, $N=20$ for $L=75$ and $N=5$ for $L=100$.
The Rayleigh law is only found for $R>R_c$
and the parameters scale to zero as power law when $R \to R_c$.
All units are arbitrary.}
\label{fig:4}
\end{figure}

\end{document}